# The Lorentz Transformation for Photons in Dispersive Media and in Gravitational Fields

ROBERT J. BUENKER[*], BERGISHE UNIVERSITAET WUPPERTAL,

*Fachbereich C-Mathematik und Naturwissenschaften, Gaussstr. 20,*

*D-42119 Wuppertal, Germany*

## Abstract

It is pointed out that the traditional explanation for the observation of a non-zero energy h υ for light in free space does not apply for the analogous situation in dispersive media. Because the speed of light u is no longer equal to c in this case, the key quantity, $\gamma = (1-u^2/c^2)^{-0.5}$, is finite as a result. Since the rest energy of photons is believed to always be equal to zero, multiplying it with γ in analogy to the usual procedure employed in the special theory of relativity (STR) does not produce a nonzero result for photons in dispersive media. The experimental evidence of the Fizeau light-drag and Cerenkov radiation phenomena indicate that the Lorentz transformation in free space is nonetheless valid for light in dispersive media. Instead, the energy and momentum of photons in transparent media can be obtained from observations of the frequency, wavelength and index of refraction of the light. A modification of the Lorentz transformation in which the observed speed of light does appear explicitly is required, however, in order to accurately predict the results of measurements made when the observer is at a different gravitational potential than the object.

## I.      Introduction

Although the relativistic theory of light in free space was fully discussed in Einstein's original work on the special theory of relativity (STR) [1], a number of questions were left open with regard to light in dispersive media. Wien [2] pointed out, for example, that the fact that the group refractive index $n_g$ can be less than unity in the neighborhood of absorption

---

lines (anomalous dispersion) implies that the speed of light can exceed c, its value in free space. If one applies the ordinary Lorentz transformation to this system, the result is a violation of causality because faster-than-c speed implies that two observers might disagree on the time-order of events under these circumstances. Sommerfeld [3] concluded that this problem does not arise in nature because even in this situation the speed of *light signals* still remains less than or equal to c. In more recent times, however, evidence has been presented [4,5] which strongly indicates that individual photons can move with speed v>c in the neighborhood of absorption lines. Nonetheless, there has been a great reluctance [4,6] to accept this result at face value because of the general belief that it violates a basic principle of STR.

There is a more general difficulty with the relativistic theory of light dispersion, however. Einstein pointed out [1] that photons were exempt from the prescription of STR that the speed of objects must always be less than the limiting value of c. He based this conclusion on the energy-momentum form of the Lorentz transformation and the fact that $\gamma = (1-u^2/c^2)^{-0.5}$ is infinite when the speed of the object equals c. The value of $\gamma$ is finite for photons in dispersive media, however, because their speed u is not equal to c in this case. The consequences of this fact have received very little attention over the years since Einstein first published his theory. At the same time, it is helpful to recall that there is another physical situation in which the speed of light is not equal to c, namely in the presence of gravitational fields. It will be seen that in the latter case a simple modification of the Lorentz transformation itself is sufficient to enable a proper description of the relevant experimental data, whereas in the case of light dispersion a different approach is required.

## II. Lorentz Transformation

The energy-momentum form of the Lorentz transformation for two inertial systems S and S' moving with relative speed u along the x direction is given below as

$$E = \gamma(u) (E' + u\, p_x') \qquad (1\text{ a})$$

$$p_x = \gamma(u) (p_x' + u\, E'/c^2), \qquad (1\text{ b})$$

with $\gamma(u) = (1 - u^2/c^2)^{-0.5}$. In free space light is observed to travel at a constant speed c ($2.997 \times 10^8$ m/s) and Einstein stated in his second postulate [1] that this value is independent of both the light source and the state of motion of the observer. Michelson and Morley [7] had earlier demonstrated that the speed of light is independent of its direction, so this aspect of Einstein's postulate had already been verified. According to Hamilton's canonical equations, the speed v of an object is related to its energy and momentum by the derivative

$$dE/dp = v. \tag{2}$$

Division of the above two equations for the case of an object traveling along the x axis leads to Einstein's velocity addition law:

$$v = (v' + u)/(1 + v'u/c^2), \tag{3}$$

where $v' = dE'/dp'$, consistent with eq. (2). Substitution of v'=c in eq. (3) leads to the result, v=c, independent of the value of the relative speed of S and S'. The same result is obtained in the more general case when the light is not moving along the direction of relative motion of the two observers (in which case Einstein assumed that $p_y=p_y'$ and $p_z=p_z'$ for the two orthogonal directions). Subsequent experiments by Kennedy and Thorndike [8] verified that the speed of light is equal to c when observations are made from the earth's surface at various times of the year. The Galilean transformation introduced by Newton holds that velocities are additive (v = v' + u), and thus could not explain either of these observations.

In applying eq. (1) to light, Einstein assumed [1] that E' = 0 and $p_{x'}$ =0, that is, that the photons at rest (u=0) have these values. This choice is sometimes seen to be at odds with the more general view that it is impossible for the observer to be in the rest frame of the photons. It has been pointed out [9], however, that because of Planck's relation,

$$E = h \nu, \tag{4}$$

photons with null energy cannot be observed because of their vanishing frequency and infinite wavelength, and thus might actually exist in great numbers in an undetectable form (the latter possibility also receives theoretical support from Bose-Einstein statistics [10]). In any event, Einstein's assumption of null values for both E' and $p_x'$ leads to a mathematically intractable result in eq. (1). Since visible light travels with speed u=c, γ=∞, whereas the linear combinations of E' and $p_x$' with which it is multiplied are equal to zero. Einstein argued successfully [1] that the physical interpretation of this result is that photons can possess non-zero energy, in accordance with eq. (4), whereas it is impossible for any other particle with nonzero rest energy E' to attain speed u=c relative to the observer because γ is not bound in this case. For such particles one has the rest condition, p' = 0, in which case eq. (1a) reduces to E= γ E', which upon squaring leads to the well known relation between E, p and E':

$$E^2 - p^2c^2 = E'^2. \qquad (5)$$

For the special case of photons (E'=0), eq. (5) reduces to

$$E = pc. \qquad (6)$$

When the latter equation is combined with Planck's energy-frequency relation in eq. (4), the result is [11]:

$$p = h \nu / c = h / \lambda, \qquad (7)$$

since in free space λ ν = c (λ is the wavelength of light).

The speed of light $c_M$ in dispersive media is generally not equal to c, but rather is determined by the group refractive index $n_g$:

$$c_M = c / n_g. \qquad (8)$$

The latter quantity is related to the refractive index n by

$$n_g = n + \nu \, dn/d\nu. \qquad (9)$$

Einstein's argument for the nonzero energy of photons in free space does not hold when $n_g$ does not equal unity because when u = $c/n_g$, γ (u) in eq. (1a) is finite in this case. If one continues to assume that both E' and p' have null values, the conclusion from eq. (1) is that E

= 0 and p = 0 for light in dispersive media. This is in contradiction to experiment, however, since one knows that the energy of photons is unchanged as it passes through a transparent medium such as water. The frequency of light is also unchanged as it passes from one medium to another, and so on this basis one is also led to conclude from eq. (4) that its energy should be unchanged as well. Since the wavelength of light is inversely proportional to n, the conclusion from eq. (7) is that the momentum of light also does not vanish in a dispersive medium, but rather increases in direct proportion to n [12, 13].

The reaction of the physics community to this state of affairs has generally been to ignore the questions that are raised by the above set of theoretical and experimental results. There has simply been a reluctance to apply STR to the phenomenon of light dispersion, given the above facts, with one clear exception. In 1907 von Laue [14] applied Einstein's velocity addition formula to the Fizeau experiment in which the "drag" effect on the speed of light in a moving medium was demonstrated [15]. By substituting v' = c /n in eq. (3), von Laue was able to obtain the observed dependence of the light speed $c_M$ on u, namely

$$c_M = c/n + u(1 - n^{-2}). \qquad (10)$$

(Here it should be noted that consistency with eq. (8) requires that n be replaced by $n_g$ in the above formula). In so doing, he showed once and for all that STR is applicable to light in dispersive media. Yet when one sets u=$c_M$ in eq. (1a) and continues to assume with Einstein that both E' and p' are equal to zero, the result is contradictory to experimental observation because it indicates that photons lose all their energy and momentum the instant they enter into a dispersive medium.

From a purely theoretical point of view, there is a simple way to remove this contradiction and still remain consistent with eq. (10). That is to assume that eq. (1) is valid when the actual observation of the interference phenomenon is made outside the medium in free space, but not within the medium itself. Instead, one can define a Lorentz transformation of the same form as eq. (1), but to substitute $c_M = c/n_g$ for c wherever it occurs:

$$E = (1 - u^2/c_M^2)^{-0.5}(E' + up_x') \qquad (11\text{ a})$$

$$p_x = (1 - u^2/c_M^2)^{-0.5}(p_x' + uE'/c_M^2). \qquad (11\text{b})$$

In this set of equations the quantity in parentheses analogous to $\gamma(u)$ in eq. (1) is infinite for $u = c_M$. Consequently, multiplying it with the null quantities on the far-right sides of eq. (11) invariably assumed for photons would then give an indeterminate, and therefore possibly non-zero result for their energy and momentum when moving at their normal speed within a given medium. It will be seen in the following discussion that eq. (11) does not give a satisfactory description of light dispersion, but that it is nonetheless relevant for gravitational interactions.

## III. The Fizeau Light-Drag Effect

Division of eq. (11a) by eq. (11b) leads to a different velocity addition law [see eq. (3)] within the medium, namely

$$v = (v' + u) / (1 + v' u / c_M^2). \qquad (12)$$

This means in effect that the speed $c_M = c/n_g$ would play the same role in the medium that the light speed c has for the Lorentz transformation in free space. As long as the light speed is v' = $c/n_g$ for the medium at rest, the measured light speed v *within the medium when it is moving* relative to the observer with speed u would always be equal to $c/n_g$ as well. In other words, the drag effect that Fizeau observed by measuring interference effects *outside the medium* would disappear if the detector is placed within the moving medium.

It is important to recall that the effect the medium has on light is quite specific. While the speed of light decreases by more than one-third when it enters water from air, there is no evidence that the speeds of other particles such as electrons are affected at all as long as they do not undergo molecular collisions within the denser medium. As a consequence there is every reason to expect that the Lorentz transformation of eq. (1) is still applicable for particles with nonzero rest energy E' when they pass through dispersive media. This assumption is necessary to explain the fact that electrons can travel faster than light in water (Cerenkov radiation). If eqs. (11a, 11b) were to be used for electrons, then it would be impossible for them to attain a speed greater than $c/n_g$, because at the moment they assume a speed equal to this value an observer moving with the electrons would necessarily find that they possess infinite energy, a totally unacceptable theoretical result, as Einstein first pointed out in his 1905 paper [1]. One might assume that a different Lorentz transformation applies for photons than for electrons in this situation because of the fact that the medium has a much different effect on the former, but this possibility seems overly complicated. A conceivable experimental test would be to place the detector within the medium in the above Fizeau experiment. If eq. (11) actually does hold for photons in dispersive media, then no drag effect should be observed under these conditions. Even if that could actually be shown, however, one would still be left with the problem of finding a consistent explanation for the fact that the speed of the electrons exceeds that of the photons regardless of where the latter are detected.

In summary, the question of why photons appear to have nonzero energy in a dispersive medium despite the fact that their speed is less than c is quite unlikely to be answered by assuming that a different form of the Lorentz transformation is applicable for light in dispersive media than in free space. The situation is different for light in a gravitational field, however, as will be discussed below.

## IV. Einstein Causality

The motivation behind introducing eq. (11) is to provide a consistent explanation as to why photons have nonzero energy even when their speed is not equal c. The key feature of the Lorentz transformation of eq. (1) and the related velocity formula of eq. (3) is that it ensures that the speed of light in free space is always equal to c no matter how fast S and S' are moving relative to one another. Clearly, by substituting a different value $c_M$ for c in these equations, as has been done in eq. (11), the only thing that changes is the value of this limiting speed. Light still travels at the same speed in all directions and this value remains independent of the state of motion of the source.

As discussed in the Introduction, one of the issues that arises in applying the Lorentz transformation is whether light can ever travel faster than c. The same objection to this possibility that Einstein made for particles of nonzero rest mass [1] does not hold for photons. Instead, emphasis has been placed on a perceived consequence of the space-time version of eq. (1):

$$dx = \gamma(u) (dx' + u\, dt') \qquad (13\text{ a})$$

$$dt = \gamma(u) (dt' + u\, dx'/c^2). \qquad (13\text{ b})$$

If one substitutes $dx' = cdt'/ n_g$ in eq. (13 b) for light in a dispersive medium, the result is

$$dt = \gamma(u)\, dt'\, (1 + u/n_g\, c). \qquad (14)$$

If $n_g$ is less than unity, as is known to occur in anomalous dispersion, then it is possible for the speed u of S' relative to S be greater than $n_g$ c. If the source moves in the opposite direction to that of the emitted light (negative u), it is mathematically possible for dt and dt' to have opposite signs, which would be a violation of Einstein causality, that is the condition that the time-order of events must be the same for all observers.

In recent work [16], however, it has been shown that such a conclusion is incorrect because it fails to take into account an ambiguity in the definition of phase relationships in the Lorentz transformation equations. To see this point, it is helpful to first consider the conventional situation in which the value of $n_g$ is greater than unity (Fig. 1). In this case c dt is always greater than dr = dx, no matter what the value of the relative speed of S and S'. Of special interest is the speed u = - c/ $n_g$. This represents the well-known situation in which two observers are temporarily traveling at the same speed (dr=0). If one of the observers increases his speed, it appears to him that dr has the opposite sign as when he decreases his speed, as indicated in Fig. 1. From the point of view of the Lorentz invariance condition, $dr^2 - c^2 dt^2 = 0$, it would be just as well to assume that dr always remains positive, however. The above observations simply show that this would be an incorrect choice of phase. In other words, the choice must be made on the basis of additional experimental information, namely that the direction of the other observer is found to change as he is passed. Another physical requirement is that both dr and dt change continuously, but this alone is of no help in fixing the phase of dr as it passes through a null value.

The situation is different when $n_g$ is less than unity, however, as illustrated in Fig. 2. In this case, cdt is always less than dr. It reaches a null value for β = u/c = -0.5 in the diagram (for $n_g$ = 0.5). Just as before (Fig. 1), one is faced with a choice of phase as the critical speed is reached. Only in this case, there is no experiment to insure that the proper assignment of phase is made. On the grounds of continuity alone, one can choose either sign for dt in Fig. 2 as β is decreased below the critical value of –0.5. Lorentz invariance is guaranteed in either case. By allowing dt to only have positive values, one avoids any violation of Einstein causality without coming into conflict with any mathematical condition required by the Lorentz invariance condition. One might argue that the fact that the speed of light becomes unbounded at the critical value of β is "unphysical," but there is also no compelling reason for ruling out this possibility because it only requires a point singularity. Accordingly, an observer would simply find that the speed of light increases without any apparent limit when β approaches the critical value in either direction, and would never actually be able to have a sustained relative speed u for which dt is exactly zero, that is, to an infinite number of significant figures.

The fact is that light speeds greater than c have been observed in free space [4, 5]. Belief in the position that such observations violate Einstein causality has led to a variety of theoretical arguments [2, 3, 6] that effectively claim that the experimental results are simply an artifact of the measurements themselves. Each of these arguments is incapable of experimental testing, however. For example, one cannot actually prove that a different part of a wave front is detected at the beginning of the measurement than at its conclusion.

## V. Inverse Lorentz Transformation

It is well known that the Lorentz transformation equations can be inverted by interchanging the primed and unprimed symbols in eqs. (1a, 1b) and changing the sign of u. It is important to see that this algebraic manipulation does not hold in the case of photons, however. Since $\gamma(u)$ is infinite in this case, it is not possible to simply divide both sides of these equations by this quantity, and that is essential in obtaining the inverted form of the transformation equations. Only when the object has nonzero rest energy E' are the inverse equations given below valid:

$$E' = \gamma(u) (E - u\, p_x) \tag{15 a}$$

$$p_x' = \gamma(u) (p_x - u\, E/c^2). \tag{15 b}$$

In the latter case, an observer traveling with the object will find that p'=0. Inserting this value into eq. (15 b) then leads directly to Einstein's famous mass-energy equivalence relation:

$$m_I = p/u = E / c^2. \tag{16}$$

Because it is not possible to effect the inversion of these equations for photons, that is, for particles with a null rest energy E', one has to find another procedure to obtain the relationship between speed and momentum in this case. As discussed elsewhere [12], this objective can be achieved by using Planck's relation of eq. (4) for the energy of photons, and then assuming that eq. (7) is also valid for light in dispersive media. Since $\lambda$ is inversely proportional to n, it then follows that the momentum of the photons is proportional to n, as Newton claimed in his original work ("Opticks"). On this basis, one is led to a generalization of eq. (6) that is applicable to the case of light in dispersive media, namely:

$$E = pc / n. \tag{17}$$

The inertial mass of the photon can then be obtained by computing the ratio p/u, whereby u is given by eq. (8). The result is different than in eq. (16), however. Instead, one obtains

$$m_I = p/u = (nE/c) / (c/n_g) = n\, n_g\, E/ c^2 = n\, n_g\, h\, \nu / c^2. \tag{18}$$

There is no contradiction with eq. (16) because the latter is not applicable to particles with null rest energy.

Finally, by way of verification of the above result, it is possible to combine eq. (2) with eq. (17) to obtain the observed dependence of the speed of light in dispersive media on the group refractive index, that is, eqs. (8, 9):

$$n_g = c/c_M = c\, dp/dE = d(nE)/dE = n + E\, dn/dE = n + \nu\, dn/d\nu, \tag{19}$$

whereby eq. (4) has again been used in the last step. Because of the fact that E' = 0 for photons, it is not possible to use the same relationships to describe them that otherwise hold quite generally for other particles. Nonetheless, the quantum mechanical relations of eqs. (4, 7) allow one to compute the energy and momentum and related properties of light from a knowledge of the frequency and wavelength of the associated light waves. The above arguments indicate that this statement holds true in all kinds of dispersive media, not just in free space.

## VI. Application to Gravitational Fields

The modification of the Lorentz transformation given in eq. (11) does not appear to apply in dispersive media, but there is another area where it is valid in free space, namely when the object of the measurement is at a different gravitational potential than the observer. Einstein pointed out [17] that because of the fact that clock rates vary with distance to a gravitational source, it follows that the speed of light is not always equal to c. As discussed in a companion article [18], the key point is that the change in the speed of light when it enters a dispersive medium from free space is an interaction between the photons and the molecules it encounters in the medium. Clearly, electrons are not subject to the same interaction, which is the main reason for the phenomenon of Cerenkov radiation.

The situation is different for the gravitational effect on light speed in free space, however. The fact that clocks slow down as they are brought closer to a gravitational source affects the measurement of the speeds of all objects. All observers, regardless of their position in a gravitational field or state of motion, must agree on the ratios of all relative speeds between two fixed points in a given inertial system [19]. Any disagreement in the absolute speed of a given object is caused purely by the fact the units of length and time are not the same for the two observers. As a consequence of this fact, eq. (11) is applicable to the motion of all objects in an area of space for which the speed of light measured by the observer has a constant value in free space of $c_M$. Indeed, eqs. (11) can be looked upon as generalization of the conventional Lorentz transformation. Altering the value of "c" in these equations simply changes the upper limit for measured speeds. This means, for example, that electrons can attain faster-than-c speeds when they are located at a higher gravitational potential than on the earth's surface, but this value must still be less than the pertinent light speed value $c_M$ in eq. (11). Moreover, when the observer changes his own position in the field, he needs to adjust the value of $c_M$ accordingly for observations at the original gravitational potential. Clearly, the equations only remain valid so long as the object stays at the same gravitational potential during the time of the measurement.

## VII. Conclusion

The fact that photons possess energy and momentum when passing through dispersive media cannot be understood on the basis of the conventional Lorentz transformation given in eqs. (1,13). Einstein's original argument that $\gamma$ is infinite while E' = 0 only holds in free space. When passing between different media photons retain their energy, so if one wants to extend Einstein's arguments beyond free space, it is necessary to either assume that E' is no longer zero for photons in dispersive media or that the value of $\gamma$ in the Lorentz transformation is still infinite despite the fact that the speed of light is not equal to c. Experimental evidence from the Fizeau light-drag and Cerenkov radiation effects speaks strongly for the conclusion that the value of $\gamma$ for light in dispersive media is simply not relevant for determining the energy of the associated photons, however. Instead, it is shown that the energy and momentum of photons in transparent media can be determined in a consistent manner by employing the well-known quantum mechanical relationships for energy/frequency and momentum/wavelength in conjunction with the known refractive index of a given medium.

The fact that the conventional Lorentz transformation retains its validity for light in transparent media is not inconsistent with Einstein causality, however. The phenomenon of anomalous dispersion indicates that the speed of light exceeds c in the neighborhood of absorption lines from the vantage point of an observer moving with high speed relative to the light source. It is shown that one can avoid a violation of Einstein causality in this case by assuming that the phase on the right-hand side of eq. (14), the relevant equation for dt in this application, changes sign as the relative velocity is varied through a critical value.

A modified version of the Lorentz transformation in which the value of c is replaced by the observed speed of light $c_M$ does find application in gravitational interactions, however. The measurement of all speeds is affected by the fact that the rates of clocks vary with their distance from a gravitational source. As a result the Lorentz transformation for an observer located at a different gravitational potential than the object of the measurement needs to employ $c_M$ rather than c to obtain satisfactory agreement with experiment. The key distinction in this case vis-à-vis observations in dispersive media is that the effect of the medium is quite different for light than for other substances, whereas gravitational changes in the units of time and distance are the same for all objects.

## Figure Captions

Fig. 1. Schematic diagram showing the variation of the spatial and time intervals dr and cdt and their ratio as a function of the relative speed $\beta = u/c$ of a light source for photons traveling at one-half the speed of light ($n_g = 2$) in free space. Note that both dr and dr/cdt vanish when

β = - 0.5 c (light travels in the opposite direction as the source) and that dr is negative for β < -c/2.

Fig. 2. Schematic diagram showing the variation of the spatial and time intervals dr and cdt and their ratio as a function of the relative speed β = u/c of a light source for photons traveling at twice the speed of light ($n_g$ = 0.5) in free space. Note that dt vanishes when β = - 0.5 c (light travels in the opposite direction as the source), causing a point singularity in the observed light speed dr/cdt. Einstein causality (dt/dt' < 0) is avoided by choosing dt >0 for β < - 0.5 c, that is, by altering the phase of the Lorentz transformation at the point of singularity and thereby still maintaining continuity for dt.

(Oct. 1 , 2005)

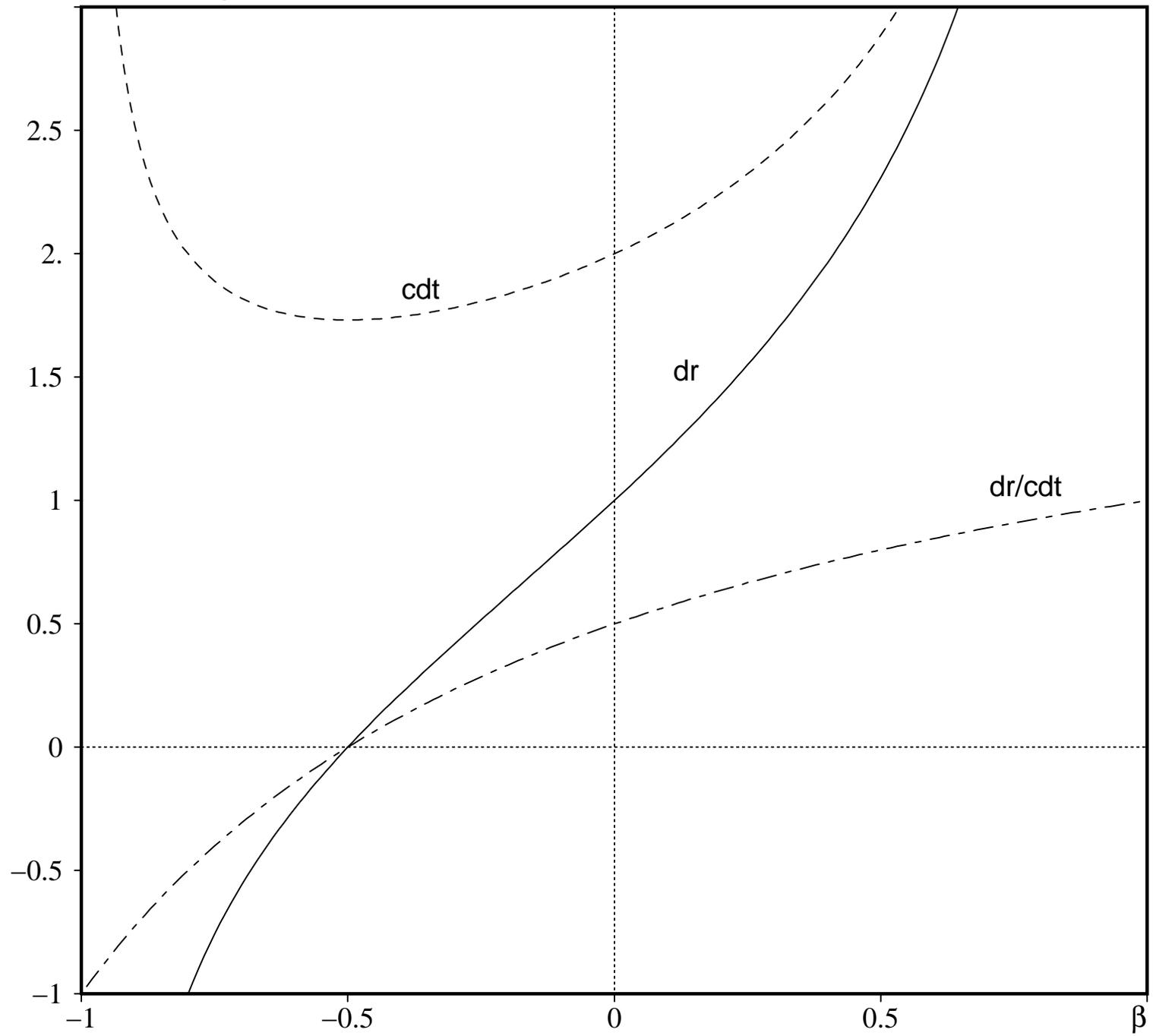

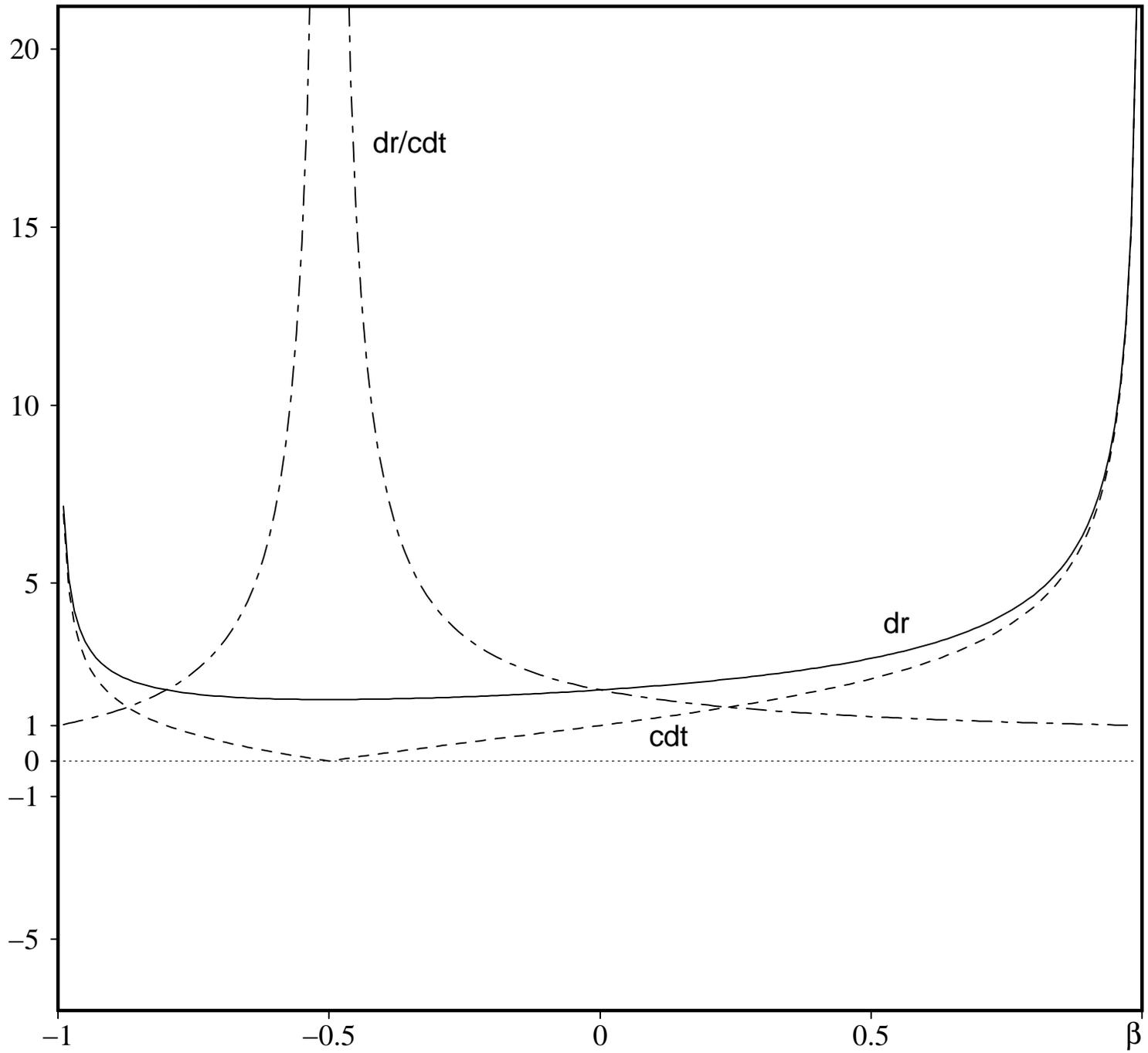